\documentclass{article}%
\usepackage{amsmath}
\usepackage{amsfonts}
\usepackage{amssymb}
\usepackage{graphicx}
\usepackage{epstopdf}
\usepackage{cite}
\usepackage[usenames]{color}%
\providecommand{\U}[1]{\protect\rule{.1in}{.1in}}
\newcommand{\ba}{\begin{array}}
\newcommand{\ea}{\end{array}}
\newcommand{\Dsl}[1] { \setbox0=\hbox{$#1$}     
\dimen0=\wd0   \setbox1=\hbox{/} \dimen1=\wd1  \ifdim\dimen0>\dimen1        
 \rlap{\hbox to \dimen0{\hfil/\hfil}}  #1 \else \rlap{\hbox to \dimen1{\hfil$#1$\hfil}}  /  \fi  }
\newcommand{\bea}{\begin{eqnarray}}
\newcommand{\eea}{\end{eqnarray}}

\newcommand{\ns}{\Dsl{n}}
\newcommand {\nbs}{\Dsl{\bar n}}

\begin{document}
\noindent

\title{ \Large  
$\eta_{c}\rightarrow  l^{+}l^{-}$ and  $\chi_{c0}\rightarrow l^{+}l^{-}$
decays revisited
}

\author{ N. Kivel$^1$\thanks{On leave of absence from St.~Petersburg Nuclear Physics
Institute, 188350, Gatchina, Russia} \  
and  A. Kupsc$^2$
\\[3mm]
{\it $^1$Helmholtz Institut Mainz, Johannes Gutenberg-Universit\"at, D-55099
Mainz, Germany}
 \\
{\it $^2$Department of Physics and Astronomy,  Uppsala University, SE-751 20 Uppsala, Sweden} 
 }

\date{}

\maketitle

\begin{abstract}
We present a calculation of $\eta_{c}\rightarrow l^{+}l^{-}$ and
$\chi_{c0}\rightarrow l^{+}l^{-}$ decay widths.  The amplitudes are
computed within leading-order approximation using NRQCD framework.
Numerical results for the branchings fractions are presented.
\end{abstract}

\bigskip

\emph{Introduction.} The leptonic decays of $C$-even charmonia have
very small branching fractions because the amplitudes are suppressed
by $\alpha^{2}$ with respect to the two photon decay modes.  For the
(pseudo-)\-scalars $\eta_{c}$ and $\chi_{c0}$ no
experimental determination of an upper limit for the dileptonic branching
fractions has been reported.  Experimental studies of 
$\eta_c$ and $\chi_{c0}$ decays usually  use the mesons
produced by radiative transitions of the vector charmonia: $J/\psi$ and
$\psi(2S)$, respectively. However, the searches for the dielectron
decay modes could instead use formation processes: $e^+e^-\to
\eta_{c}$ and $e^+e^-\to \chi_{c0}$ where the (pseudo-)\-scalar meson
production is tagged using one of its common decays. This method has
an advantage of low background and could provide high sensitivity.
The method has been applied {\it e.g.}  for searches of the $\eta'\to
e^+e^-$ process at VEPP-2000 where an impressive upper limit for the
branching fraction of $5.6\times 10^{-9}$ at 90\% C.L.  was achieved
using integrated luminosity of $2.9$ pb$^{-1}$
\cite{Akhmetshin:2014hxv,Achasov:2015mek}. In case of $C$-even
charmonia such experiments are possible at BEPC-II collider with the
BESIII detector \cite{Asner:2008nq}.

\begin{figure}[h]%
\centering
\includegraphics[width=4.0in]%
{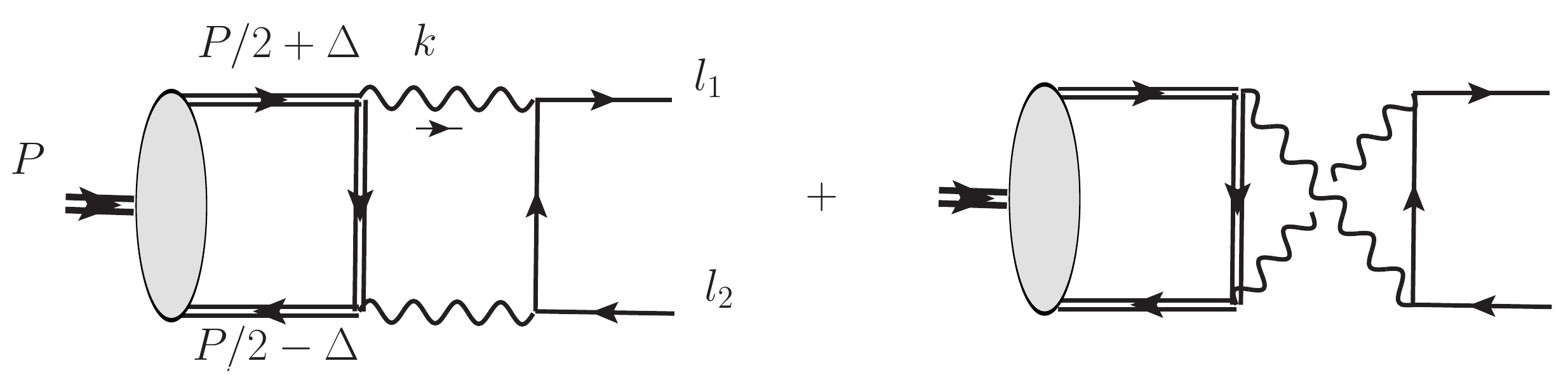}%
\caption{One-loop diagrams describing the annihilation into lepton pair with
momenta $l_{1}$ and $l_{2}$.}%
\label{diagrams}%
\end{figure}

Calculations of the leptonic decay amplitudes can be carried out using
the NRQCD framework, see {\it e.g.}
Refs.\cite{Bodwin:1994jh,Beneke:1997zp,Brambilla:2004jw}.  Recently
such  calculations have been performed for the $\chi_{c1}$ and
$\chi_{c2}$ decays in Ref.\cite{Kivel:2015iea}.  The decays
$\eta_{c}\rightarrow l^{+}l^{-}$ and $\chi _{c0}\rightarrow
l^{+}l^{-}$ can also be described in the same framework. However, the
corresponding amplitudes are suppressed by an additional factor
$m_{l}/m_{c}$ where $m_{l}$ and $m_{c}$ are lepton and charm quark
masses, respectively.  This is a consequence of the conservation of the
orbital momentum: the lepton helicity flip is mandatory in decays of
(pseudo-)scalar mesons.  The dominant diagrams with two photons in the
intermediate state are shown in Fig.\ref{diagrams}.
The gray blob in the figure denotes the charmonium bound state with
momentum $P$.  In this figure we assume that the dominant
contribution is associated with the leading-order $Q\bar{Q}$ component
of the wave function.  This assumption is valid if the dominant
contribution to the corresponding loop integral comes from region(s) with
the large virtuality of the intermediate heavy quark $\left(
\frac{1}{2}P+\Delta-k\right) ^{2}-m^{2}\gg(mv)^{2}$, where $m$ denotes
the heavy quark mass and $v$ is the small relative velocity of the
heavy quarks. On the other hand the virtualities of the photons and
lepton can be arbitrary because these particles belong to the QED
sector. For such case the resulting integral yields
the leading-order approximation and the overlap with the physical
state is described by the matrix element which can be associated with
the two quark component of the charmonium wave function. However, as it
was shown in Ref.\cite{Kivel:2015iea} such simple picture is not valid
for the $P$-states and resulting interpretation is more complicated.
In the following we provide a short description for the decay
amplitudes of $\eta_c$ and $\chi_{c0}$.

\emph{Calculation of the amplitude and branching fraction for} $\eta
_{c}\rightarrow l^{+}l^{-}$\emph{.} The decay amplitude reads
\begin{equation}
A_{\eta_{c}\rightarrow ll}\simeq D_{\gamma\gamma}~\left\langle 0\right\vert
~\psi_{\omega}^{\dag}\gamma_{5}\chi_{\omega}~\left\vert \eta_{c}\right\rangle
,\label{Aeta}%
\end{equation}
where $\psi_{\omega}^{\dag}$ and$~\chi_{\omega}$ denote the heavy quark fields
in the heavy quark effective theory (HQET), $\omega$ is the velocity of the
heavy meson%
\begin{equation}
P=M\omega=l_{1}+l_{2},
\end{equation}
$M$ denotes the mass of $\eta_{c}$ and  we use the  rest frame
where $\omega=(1,\vec{0})$. The HQET fields satisfy
\begin{equation}
\psi_{\omega}^{\dag}\Dsl{\omega}=\psi_{\omega}^{\dag},~\ \Dsl{\omega}
\chi_{\omega}=-\chi_{\omega}.
\end{equation}
The matrix element in Eq.(\ref{Aeta}) reads
\begin{equation}
\left\langle 0\right\vert ~\psi_{\omega}^{\dag}\gamma_{5}\chi_{\omega
}~\left\vert \eta_{c}\right\rangle =\sqrt{2M}\sqrt{\frac{3}{2\pi}%
}~R_{10}(0),
\end{equation}
where $R_{10}(0)$ is the radial component of the charmonium wave function at
the origin. 

The  coefficient $D_{\gamma\gamma}$ in Eq.(\ref{Aeta}) is given by the
diagrams in Fig.\ref{diagrams}  and reads (Feynman gauge, $e_{c}=2/3$, is used)%
\begin{align}
D_{\gamma\gamma}  & =-\alpha^{2}e_{c}^{2}\int\frac{d^{4}k}{i\pi^{2}}~\bar
{u}(l_{1})D_{l}^{\alpha\beta}v(l_{2})\frac{1}{\left[  k^{2}-2m(k\omega
)\right]  }\nonumber\\
& ~\ \ \ \ \ \times~\frac{1}{4}\text{Tr}\left[  (1+\Dsl{\omega})\gamma
_{5}\gamma^{\beta}(m\Dsl \omega-\Dsl k+m)\gamma^{\alpha}+\gamma^{\alpha
}(\Dsl k-m\Dsl \omega+m)\gamma^{\beta}\right]  .\label{Dgg-eta}%
\end{align}
In the above expression the small relative momentum in the heavy quark
propagator is neglected%
\begin{equation}
\left(  \frac{1}{2}P+\Delta-k\right)  ^{2}-m^{2}\simeq k^{2}-2m(k\omega
).\label{HQprop}%
\end{equation}

In the numerator of the leptonic part we keep the linear terms in $m_{l}$
\begin{equation}
\bar{u}(l_{1})D_{l}^{\alpha\beta}v(l_{2})\simeq\bar{u}_{n}\left(  1+\frac
{\nbs}{2}\frac{m_{l}}{M}\right)  \frac{\gamma^{\alpha}(\Dsl{l}_{1}-\Dsl k+m_{l}%
)\gamma^{\beta}}{\left[  k^{2}\right]  \left[  \left( k-l_{1}\right)
^{2}-m_{l}^{2}\right]  \left[  \left(  k-P\right)  ^{2}\right]  }\left(
1-\frac{m_{l}}{M}\frac{\ns}{2}\right)  v_{\bar{n}}.\label{leptNum}%
\end{equation}
The last  expression uses auxiliary light cone vectors $n$ and $\bar{n}$
related to the lepton momenta%
\begin{equation}
l_{1}=M\frac{n}{2}+\mathcal{O}(1/m),~\ \ l_{2}=M\frac{\bar{n}}{2}%
+\mathcal{O}(1/m).
\end{equation}
The spinors in Eq.(\ref{leptNum}) has been decomposed as%
\begin{equation}
\bar{u}(l_{1})\simeq\bar{u}_{n}\left(  1+\frac{\nbs}{2}\frac{m_{l}}%
{M}\right)  ,~\bar{u}_{n}=\bar{u}(l_{1})\frac{\nbs\ns}{4},\
\end{equation}
\begin{equation}
v(l_{2})\simeq\left(  1-\frac{m_{l}}{M}\frac{\ns}{2}\right)  v_{\bar{n}%
},~\ v_{\bar{n}}=\frac{\nbs\ns}{4}v(l_{2}).
\end{equation}

Using  the threshold expansion technique developed in Ref.\cite{Beneke:1997zp}, one  finds the following
dominant regions: hard $k\sim m$, lepton collinear~\ $k\sim l_{1}$ or $k\sim
l_{2}$ and  lepton ultrasoft  $k-l_{1}\sim m_{l}$ with $\left(  k-l_{1}%
\right)  ^{2}-m_{l}^{2}\sim m_{l}^{2}$.  In  all cases the virtuality  of
the heavy quark propagator  is of order $m^{2}$, {\it i.e.} large. We can
therefore proceed with the loop calculations neglecting the small momentum components as
it is done in Eq.(\ref{HQprop}).  The lepton mass can not be completely
neglected because it serves as a natural regulator in the collinear and
ultrasoft regions.  Therefore the result depends on the large logarithms
$\ln m_{l}/m_{c}$.  Computing the integral in Eq.(\ref{Dgg-eta}) we obtain%
\begin{equation}
A_{\eta_{c}\rightarrow ll}=\left\langle 0\right\vert ~\psi_{\omega}^{\dag
}\gamma_{5}\chi_{\omega}~\left\vert \eta_{c}\right\rangle ~\bar{u}_{n}%
\gamma_{5}v_{\bar{n}}~\alpha^{2}\frac{m_{l}}{M}\frac{e_{c}^{2}}{m^{2}}\left(
\frac{1}{4}\ln^{2}\lambda-\ln\lambda+4\ln2+\frac{\pi^{2}}{12}+\frac{i\pi}%
{2}\ln\lambda\right)  ,
\end{equation}
where
\begin{equation}
\lambda=\frac{m_{l}^{2}}{4m^{2}}.\label{def:lambda}%
\end{equation}

In order to get the numerical estimate we use $\alpha=1/137$,  $m_{c}=1.5$~GeV,
$m_{e}=0.51$~MeV, $m_{\mu}=105.6$~MeV and  the value of $R_{10}(0)$  from
Ref.\cite{Eichten:1995ch} for Buchm\"uller-Tye potential \cite{Buchmuller:1980su}:
\begin{equation}
\left\vert R_{10}(0)\right\vert ^{2}\simeq0.81~\text{GeV}^{3}.
\end{equation}
With these parameters we get
\begin{equation}
~\ Br\left[  \eta_{c}\rightarrow e^{+}e^{-}\right]  =5.6\times 10^{-13}
,\ ~Br\left[  \eta_{c}\rightarrow\mu^{+}\mu^{-}\right]  =1.66\times
10^{-9}.\label{Br-BT}
\end{equation}
Alternatively one can consider the branching fractions ratio  where $R_{10}(0)$
cancels
\begin{equation}
\frac{Br\left[  \eta_{c}\rightarrow e^{+}e^{-}\right]  }{Br\left[  \eta
_{c}\rightarrow\gamma\gamma\right]  }=1.6\times10^{-9},~\ \frac{Br\left[
\eta_{c}\rightarrow\mu^{+}\mu^{-}\right]  }{Br\left[  \eta_{c}\rightarrow
\gamma\gamma\right]  }=4.7\times10^{-6}.
\end{equation}
Using  value for $Br\left[  \eta_{c}\rightarrow\gamma
\gamma\right]  =1.57\times10^{-4}$   \cite{Agashe:2014kda}  we obtain
\begin{equation}
\ Br\left[  \eta_{c}\rightarrow e^{+}e^{-}\right]  =2.5~\times10^{-13}%
,\ ~Br\left[  \eta_{c}\rightarrow\mu^{+}\mu^{-}\right]  =0.74~\times10^{-9}.
\end{equation}
These estimates are about factor two smaller then the values in Eq.(\ref{Br-BT}). The difference can be considered as  an estimate of theoretical uncertainty of in this approach. 

The $\eta_{c}\rightarrow l^{+}l^{-}$ process has been previously
studied in Ref.\cite{Yang:2009kq} using a different theoretical
approach.  Our estimate for $Br\left[ \eta_{c}\rightarrow
  \mu^{+}\mu^{-}\right]$ is in agreement with the one from this reference
 within the uncertainties, but the results for $Br\left[
  \eta_{c}\rightarrow e^{+}e^{-}\right]$ differ by factor of six.
 
\emph{Calculation of the amplitude and branching fraction for}  $\chi_{c0}\rightarrow l^{+}l^{-}$.  
In this case the the description of
the  amplitude in the effective theory framework  is  more complicated. The
integral originating from the diagram in Fig.\ref{diagrams} has an infrared
singularity because there is a region of the integration where the heavy quark
propagator becomes soft. Therefore  in order to obtain a consistent
description in NRQCD  one has to include a contribution associated with higher
Fock  component of the charmonium wave function \ $\left\vert QQ\gamma
\right\rangle $.  This can be done in the same way as  for decay $\chi
_{cJ}\rightarrow l^{+}l^{-}$ with $J=1,2$, see {\it e.g.} Ref.\cite{Kivel:2015iea}.  In addition one has to take into account the collinear and soft regions which
could also be relevant.  Therefore the expression for the amplitude
can be represented as  a sum of two terms 
\begin{equation}
A_{\chi_{0}\rightarrow ll}\simeq i\bar{u}_{n}v_{\bar{n}}\frac{m_{l}}%
{m}~\left\{  C_{\gamma\gamma}^{(0)}(\mu_{F})\left\langle \mathcal{O}(^{3}%
P_{0})\right\rangle -\frac{\alpha}{\pi}e_{Q}C_{\gamma}\frac{1}{\sqrt{3}}%
h(\mu_{F})\right\}  .\label{Achi}%
\end{equation}
The first term in Eq.(\ref{Achi}) describes the contribution which overlaps with
$Q\bar{Q}$ components of the charmonium wave function. In this case  
\begin{equation}
\left\langle \mathcal{O}(^{3}P_{0})\right\rangle \equiv\left\langle
0\right\vert \frac{1}{2\sqrt{3}}~\chi_{\omega}^{\dag}\overleftrightarrow{D}%
_{\top}^{\alpha}\gamma_{\top}^{\alpha}\psi_{\omega}~\left\vert \chi
_{c0}\right\rangle =\sqrt{2N_{c}}\sqrt{2M_{\chi_{c0}}}\sqrt{\frac{3}{4\pi}%
}R_{21}^{\prime}(0),
\end{equation}
where $R_{21}^{\prime}(0)$ denotes the derivative of the wave function at the
origin. The subscript $_{\top}$ is used for the Lorentz indices which are orthogonal
to the velocity $\omega$, for instance, $\omega_{\alpha}\gamma_{\top}^{\alpha
}=0$.  The hard  coefficient function $C_{\gamma\gamma}^{(0)}(\mu_{F})$
is associated with the integration regions  where the heavy quark propagator is hard.  
In this case we find the same dominant regions as described above for the $\eta_{c}$ decay. 

 However in present case  there is an additional  
domain  when the photon momentum is ultrasoft, $k_{\mu}\sim mv^{2}$.
 The overlap  of the hard and the ultrasoft  regions  leads to  the logarithmic divergence that  introduces a
dependence on the factorization  scale $\mu_{F}$. In the ultrasoft  region  the heavy
quark propagator is soft  $\left(  \frac{1}{2}P+\Delta-k\right)  ^{2}
-m^{2}\sim(mv)^{2}$  and therefore the corresponding contribution  cannot be
given by the matrix element associated with $Q\bar{Q}$ component of the
charmonium wave function.  Corresponding   contribution is given by the
second term on \textit{r.h.s.} of Eq.(\ref{Achi}) where the quantity $h(\mu_{F})$
is defined  by the following  matrix element
\begin{equation}
\left\langle 0\right\vert ~\chi_{\omega}^{\dag}\gamma_{\top}^{\sigma}%
\psi_{\omega}Y_{n}^{\dag}Y_{\bar{n}}~\left\vert \chi_{c0}\right\rangle
=-\frac{1}{2}(n-\bar{n})^{\sigma}~i\frac{\alpha}{\pi}e_{Q}\frac{1}{\sqrt{3}%
}h(\mu_{F}),\label{def:h}%
\end{equation}
with the ultrasoft photon Wilson lines
\begin{equation}
Y_{n}^{\dag}=\text{Pexp}\left\{  ie\int_{0}^{\infty}ds~n\cdot B^{us}%
(sn)\right\}  ,~\ Y_{\bar{n}}=\text{\={P}exp}\left\{  -ie\int_{0}^{\infty
}ds~\bar{n}\cdot B^{us}(s\bar{n})\right\}  ,~\
\end{equation}
where $B_{\mu}^{us}$ denotes the
ultrasoft photon field.  In the leading-order approximation with respect to the
electromagnetic coupling $e$ these Wilson lines are equal to unity
$Y_{n}^{\dag}=Y_{\bar{n}}=1+\mathcal{O}(e)$. Then the  matrix element in 
Eq.(\ref{def:h}) vanishes because of $C$-parity. One has to pick up at least one term
$\sim eB^{us}$ in the expansion of the Wilson lines in order to get  the
$C$-even operator. Therefore we can conclude that the matrix element in
 Eq.(\ref{def:h}) can be associated with  the coupling to the higher Fock component  $\left\vert
Q\bar{Q}\gamma\right\rangle $ of the charmonium wave function.  The value of
the corresponding constant  $h(\mu_{F})$ in Eq.(\ref{def:h}) is the same for all states $\chi_{cJ}$ due to the
heavy quark spin symmetry. At low normalization point $\mu_{F}=\mu_{0}\simeq400$~MeV 
it can be computed in the low energy effective theory describing   interaction of the ultrasoft photons with heavy mesons.
The ultasoft matrix element in Eq.(\ref{def:h}) also contributes to the decays $\chi_{c1,1}\rightarrow e^{+}e^{-}$ and has been 
already computed  in Ref.\cite{Kivel:2015iea}
\begin{align}
h(\mu_{0})  & =f_{\gamma}\sqrt{2M_{J/\psi}}\sqrt{\frac{3}{2\pi}}R_{10}%
(0)\frac{~\Delta}{M}\left(  1-\ln2+\ln[\mu_{0}/\Delta]+i\pi \right)  \\
& +f_{\gamma}^{\prime}\sqrt{2M_{\psi^{\prime}}}\sqrt{\frac{3}{2\pi}}%
R_{20}(0)\frac{\Delta^{\prime}}{M_{\chi_{0}}}\left(  1-\ln2+\ln[\mu_{0}/|\Delta
^{^{\prime}}|]\right)  ,
\end{align}
where $\Delta=(M_{\chi_{0}}^{2}-M_{J/\psi}^{2})/2M$,~$\Delta^{\prime}=(M_{\chi_{0}}^{2}
-M_{\psi^{\prime}}^{2})/2M_{\chi_{0}}$,$~R_{10}(0)$ and$~R_{20}(0)$ denote the radial
wave functions of $J/\psi$ and $\psi^{\prime}$ mesons, respectively.  In what follows we take
their values from Ref.\cite{Eichten:1995ch} for Buchm\"uller-Tye potential. The dimensionless
couplings $f_{\gamma}$ and $f_{\gamma}^{\prime}$ can be determined from the
decays $\chi_{cJ}\rightarrow J/\psi+\gamma$ and $\psi^{\prime}\rightarrow
\chi_{cJ}+\gamma$, respectively   
\begin{equation}
f_{\gamma}\simeq6.0,~\ f_{\gamma}^{\prime}\simeq-7.2.
\end{equation}

The hard coefficient $C_{\gamma}$ in Eq.(\ref{Achi}) is given by the tree diagram
describing  annihilation subprocess $c\bar{c}\rightarrow e^{+}e^{-}$ and reads
\begin{equation}
C_{\gamma}=\frac{\alpha\pi}{m^{2}}e_{c}.
\end{equation}
The determination of the second hard coefficient
$C_{\gamma\gamma }^{(0)}$  in Eq.(\ref{Achi}) requires calculation of
the diagrams in Fig.\ref{diagrams} and one-loop calculation of the matrix
element in Eq.(\ref{def:h}) in the potential NRQED \cite{Pineda:1997bj,Brambilla:1999qa,Brambilla:1999xf}. These calculations
are similar to the ones carried out in Ref.\cite{Kivel:2015iea}.  The
only difference is that a minimal dependence on the lepton mass $m_{l}$
has to be included in order to avoid IR-singularities in the QED sector.  The final result reads
\begin{equation}
C_{\gamma\gamma}^{(0)}(\mu_{F})=\frac{\alpha^{2}}{\sqrt{3}}\frac{e_{c}^{2}%
}{m^{3}}\left\{  2\ln\frac{m^{2}}{\mu_{F}^{2}}+\frac{3}{4}\ln^{2}\lambda
+\ln\lambda+\frac{\pi^{2}}{4}+2+6\ln2+i\pi\left(  2\ln2-1+\frac{3}{2}%
\ln\lambda\right)  \right\}  ,\label{C0gg}%
\end{equation}
where $\lambda$ is defined in Eq.(\ref{def:lambda}).  
The numerical estimates of the branching fractions are
\begin{equation}
Br[\chi_{c0}\rightarrow e^{-}e^{+}]=1.0\times10^{-12},~\ Br[\chi
_{c0}\rightarrow\mu^{-}\mu^{+}]=2.2\times10^{-9} .
\end{equation}
We observe that the hard contribution with
$C_{\gamma\gamma}^{(0)}$ dominates and practically saturates the numerical
values contrary to  $\chi_{c1,2}\rightarrow l^{+}l^{-}$ decays
 where the ultrasoft contribution is the most important, see Ref.\cite{Kivel:2015iea}.  This is explained by a relative enhancement of
$C_{\gamma\gamma}^{(0)}$ in Eq.(\ref{C0gg}) by the large
logarithms $\ln\lambda$  with respect to the ultrasoft term $h$ in Eq.(\ref{Achi})
which remains unchanged. 

 The amplitude of the decay  $\chi_{cJ}\rightarrow l^{+}l^{-}$ has also been considered in Ref.\cite{Yang:2012gk} where only the 
 hard  contribution with $C_{\gamma\gamma}^{(0)}$ has been taken into account.   The result in Eq.(\ref{C0gg}) differs from the one in Ref.\cite{Yang:2012gk}  only by
simple non-logarithmic terms. We observe that this discrepancy  does not provide any tangible numerical effect.  
  The estimate for $Br[\chi_{c0}\rightarrow l^{+}l^{-}]$ obtained in this work  is about factor 
 three larger which is explained by the different choice of the numerical values used for $m_{c}$ and $R'_{21}(0)$.

\end{document}